\documentclass{ws-procs9x6}

 \def\g2{GeV$^{2}$}

\begin{document}
 
\title{Study of narrow baryonic pentaquark candidates with the
ZEUS detector at                                 
HERA\footnote{\uppercase{T}his work is supported by the \uppercase{I}srael 
\uppercase{S}cience \uppercase{F}oundation and the \uppercase{U.S.-I}srael        
\uppercase{B}i-national \uppercase{S}cience \uppercase{F}oundation}}      
 
\author{Uri Karshon\footnote{\uppercase{O}n behalf of the
\uppercase{ZEUS} \uppercase{C}ollaboration }}
 
\address{Weizmann Institute of Science, \\
Rehovot, Israel\\ 
E-mail: uri.karshon@weizmann.ac.il}

\maketitle
 
\abstracts{
The three         pentaquark candidates $\Theta^+ (1530)$,
$\Xi (1862)$ and $\Theta_c^0 (3100)$ have been studied in $ep$ collisions
at a centre-of-mass energy $\sqrt{s} = 300-318$~GeV using the full luminosity 
of the HERA-I data. Searches for narrow baryonic states in the decay
channels $K^0_S p$, $K^+ p$, $\Xi^-\pi^{\pm}$, $\bar\Xi^+\pi^{\pm}$ and
$D^{*\pm}p^{\mp}$ are reported. The results support the existence of 
a narrow resonance                       decaying into 
 $K^0_S p$ and $K^0_S\bar p$, consistent with the $\Theta (1530)$
state. No signals are seen in the
 $K^+ p$, $\Xi^-\pi^{\pm}$, $\bar\Xi^+\pi^{\pm}$ and $D^{*\pm}p^{\mp}$
channels.}

\section{Introduction and Experiment         }
 
The HERA e-p collider accelerates electrons (or positrons)
          and protons to energies of $E_e =27.5$~GeV
and $E_p =920$~GeV ($820$~GeV           
             until 1997), respectively.                                       
                                              The two collider  experiments, H1
                                                    and ZEUS,            are 
located at two collision points along the circulating beams. 
                                                             The incoming $e^{\pm}$
interacts with the proton by first radiating a virtual photon. The photon is either
quasi-real with                                          
 $Q^2 < 1$~GeV$^2$ and           
   $ Q^2_{median} \approx 3\cdot 10^{-4}$~GeV$^2$,        
                                      where $Q^2$ is the negative
 squared four-momentum transferred between the electron and proton,
                                                        or highly  virtual
($Q^2 > 1$~GeV$^2$). In the former case no scattered electron is visible and this
is the photoproduction (PHP) regime. In the latter case the scattered electron is     
measured in the main detector and this is the deep inelastic scattering (DIS) regime.
The analysis was performed with ZEUS data taken between 1995 - 2000 (``HERA-I"),
corresponding to an integrated luminosity of $\approx 120$~pb$^{-1}$.
Charged particles are tracked in the Central Tracking Detector (CTD)          
                                           covering polar angles of
$15^{\circ}<\theta<164^{\circ}$.                                                     
The energy loss of particles in the CTD, $dE/dx$, is estimated from the
truncated mean of the anode-wire pulse heights, after removing the lowest
$10\%$ and at least the highest $30\%$ depending on the number of saturated hits.
The $dE/dx$ resolution for full-length tracks is about $9\%$.
 
 \vspace*{-0.2cm}
\section{Evidence for the strange pentaquark $\Theta^+ (1530)$}
 
Fixed-target low-energy experiments saw a narrow exotic baryon with strangeness
$+1$ around $1530$~MeV decaying into $K^+ n$. It was attributed to the $\Theta^+=uudd\bar s$
pentaquark candidate predicted by Diakonov et al.~\cite{Diakonov} at the top of
a $SU(3)$ spin $1/2$ anti-decuplet of baryons. Narrow peaks were also seen at a similar mass
in the final state $K^0_S p$, which is not necessarily exotic. They were
attributed to the $\Theta^+$ as well. It is interesting to search for the
$\Theta^+$ baryon in high-energy collider  experiments. In particular it can be
searched at the central rapidity region, which has little sensitivity  to the
proton remnant region. This region is dominated by parton fragmentation with no net
baryon number, unlike low-energy experiments, where the pentaquark is mainly      
produced in the nucleon fragmentation region.
 
The ZEUS search for the $\Theta^+(1530)$~\cite{zeustheta} used the 1996 - 2000 HERA
           data ($121$~pb$^{-1}$)                                              
and was performed in the DIS regime ($Q^2 > 1$~GeV$^2$).
The search in the $K^0 p$ mode was complicated due to       a few unestablished
resonances, such as $\Sigma (1480)$ and $\Sigma (1560)$,                  called
``$\Sigma$ bumps"~\cite{PDG}. There are no such known bumps
around the $\Theta^+$ mass range; however, it is difficult to describe the background
under a $\Theta$ signal due to these 
$\Sigma$ bumps.                                          
 
$K^0_S$ particles were reconstructed from secondary-vertex CTD tracks
with transverse momenta $p_T > 0.15$~GeV and pseudo-rapidities $|\eta |~<~1.75$.
                                     The $K^0$ transverse momenta and pseudorapidities
were required to have $p_T (K^0) > 0.3$~GeV and $|\eta (K^0)| < 1.5$. A very clean
$K^0\to\pi^+\pi^-$ signal was obtained~\cite{zeustheta}.                                                 
      After requiring       
$0.483 < M(\pi^+\pi^-) < 0.513$~GeV, the number of $K^0_S$ candidates
               was $\approx 867,000$ with only $\approx 6\%$ background. 
 
 
 
 
 
 
 
 
 

 
 Protons and antiprotons                                                  
 were selected from                        a wide $dE/dx$ proton band, motivated by
the Bethe-Bloch                 equation,     defined for primary-vertex         
       tracks~\cite{zeustheta}. Pion and kaon
    contamination                    was         reduced by requiring the proton
momentum to be less than $1.5$~GeV and $dE/dx$ to be above 1.15 minimum ionising  
particles (mips). The purity of the proton sample obtained was            
$\approx 60\%$.
 
\begin{figure}[ht]
 
 
 \hspace*{-0.6cm}      
  \resizebox{15.5pc}{!}{\includegraphics{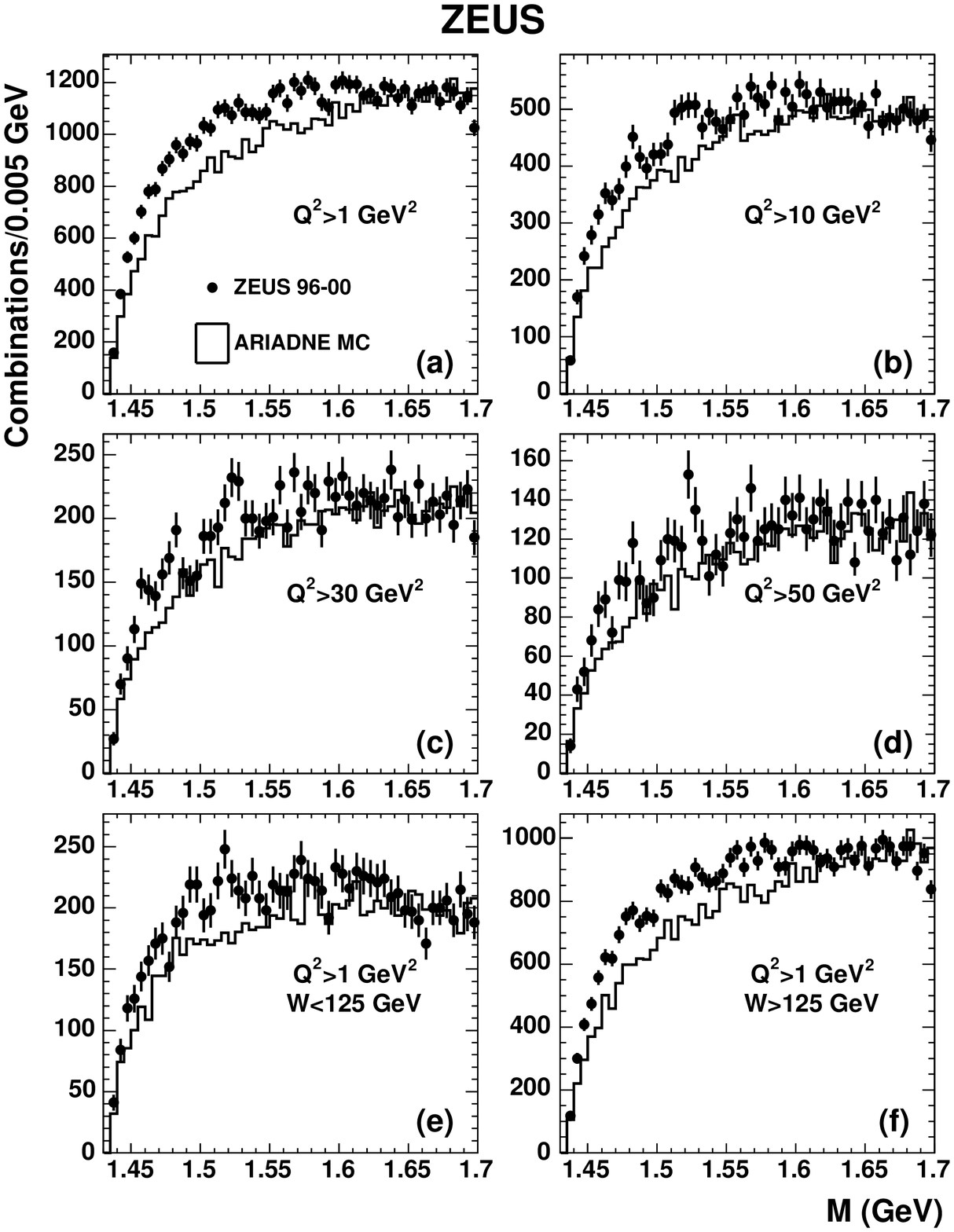}}
 
 
\vspace*{-7.8cm}\hspace*{+5.8cm}\resizebox{14pc}{!}{\includegraphics{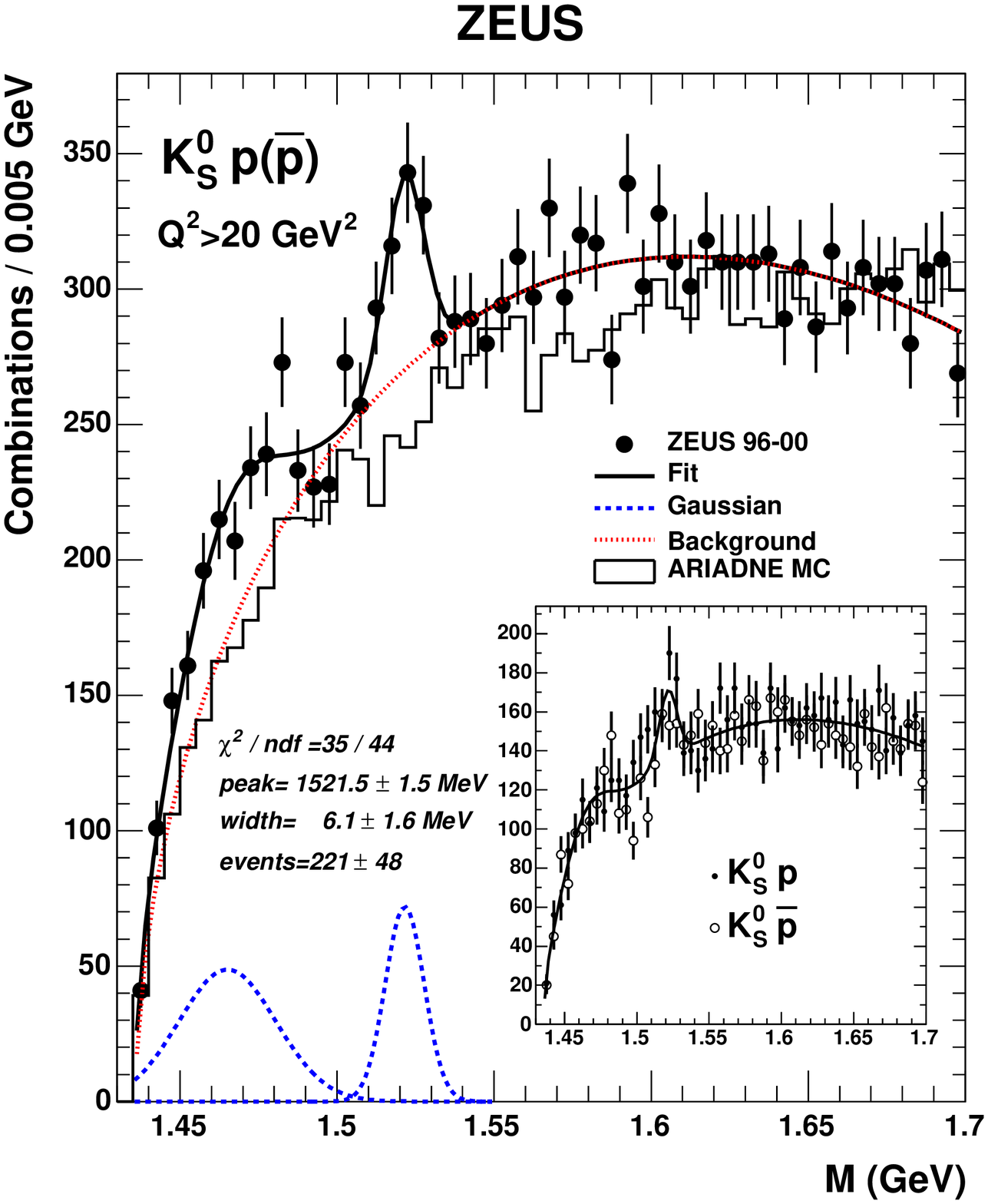}} 
 
 \vspace*{ -6.5cm}\hspace*{10.8cm}{        (g)}
 \vspace*{ +6.5cm}\hspace*{-10.8cm}               
 
 
 \vspace*{+0.1cm}                      
\caption{$M(K^0_S p (\bar p))$ for (a-d) $Q^2 > 1,10,30,50$~GeV$^2$; (e-f)    
$W < 125$ and $ > 125$~GeV; (g)     $Q^2 > 20$~GeV$^2$.                         
 The histograms are     MC predictions normalised to the data above
 $1.65$~GeV. The solid line in (g)            is a fit to the data using a background
function (dotted line) plus two Gaussians (dashed lines). The inset shows  the
$K^0_S\bar p$ (open circles) and $K^0_S p$ (black dots) combinations, compared to the
combined sample fit scaled by 0.5.
            \label{Fig2}}
\end{figure}

The $K^0_S p (\bar p)$ mass spectrum is shown in Fig.~\ref{Fig2}(a-f)            
                                with a minimum $Q^2$ ranging from $1$ to $50$~GeV$^2$,
as well as with $Q^2 > 1$~GeV$^2$ for two separate bins of the                      
                                         photon-proton centre-of-mass energy, $W$.
For $Q^2 > 10$~GeV$^2$ or $Q^2 > 1$~GeV$^2$ and $W < 125$~GeV, a peak is seen near $1.52$~GeV.
The histograms are the ARIADNE Monte Carlo (MC) simulation, normalised to the data
above $1.65$~GeV. The shape of the data distributions is not well described by the 
MC which does not simulate the
 $\Sigma$ bumps.                                          
 
 \vspace*{ -0.5cm}                                
 
 \vspace*{-0.5cm}                      
In Fig.~\ref{Fig2}(g) the 
    $K^0_S~p~(\bar p)$ mass spectrum is shown for $Q^2~>~20$~GeV$^2$ together     with
a fit to two     Gaussians     and a background     of the form \\  
$P_1(M-m_p-m_{K^0})^{P_2}\cdot(1+P_3(M-m_p-m_{K^0}))$, where $M$ is the
$K^0_S p$ mass, $m_p$    ($m_{K^0}$) is the proton ($K^0$) mass
and $P_{1,2,3}$ are free parameters.                                                 
                           The fit       $\chi^2 /ndf$ ($35/44$)    
is significantly better than a one-Gaussian fit for the $\Theta$ only. The improvement
is mainly in the low mass region, where the second resonance may correspond to the
 $\Sigma (1480)$. The $\Theta$ peak position is                        
   $M = 1521.5\pm 1.5(stat.)^{+2.8}_{-1.7}(syst.)$~MeV, with a Gaussian width
   $\sigma = 6.1\pm 1.6(stat.)^{+2.0}_{-1.4}(syst.)$~MeV, which is above, but
consistent with the resolution ($\approx 2$~MeV). The fit gives $221\pm 48$ events
above     background, corresponding to $4.6\sigma$.
The probability of a fluctuation leading to the observed signal in the mass range
$1.5 - 1.56$~GeV is below $6\cdot 10^{-5}$. Fitting the $\Theta$ with a Breit-Wigner
         convoluted with a Gaussian fixed to the experimental resolution, the
intrinsic full width of the signal is estimated to be $\Gamma = 8\pm 4(stat.)$~MeV.
The signal is seen for both proton charges (inset in 
   Fig.~\ref{Fig2}g). The fitted number of events in the $K^0_S\bar p$ channel is
$96\pm 34$. If the signal originates from the $\Theta$, this is a   first evidence
for the anti-pentaquark $\Theta^-$.
 
\begin{figure}[ht]
 
 \vspace*{-0.5cm}      
 \hspace*{-0.5cm}      
  \resizebox{15.0pc}{!}{\includegraphics{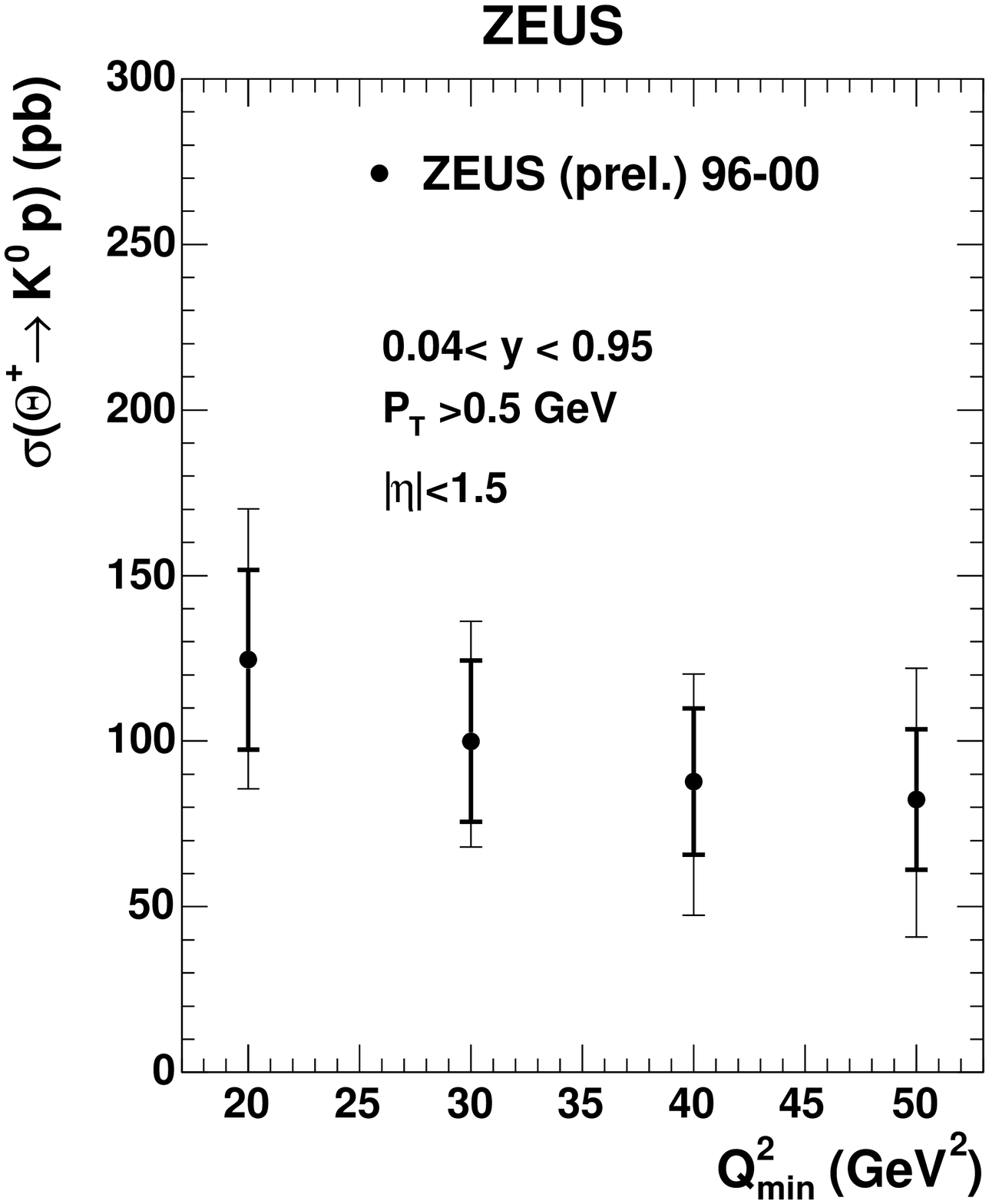}}
 
 \vspace*{ -6.5cm}\hspace*{ 0.8cm}{        (a)}
 
\vspace*{-1.4cm}\hspace*{+5.8cm}\resizebox{15.0pc}{!}{\includegraphics{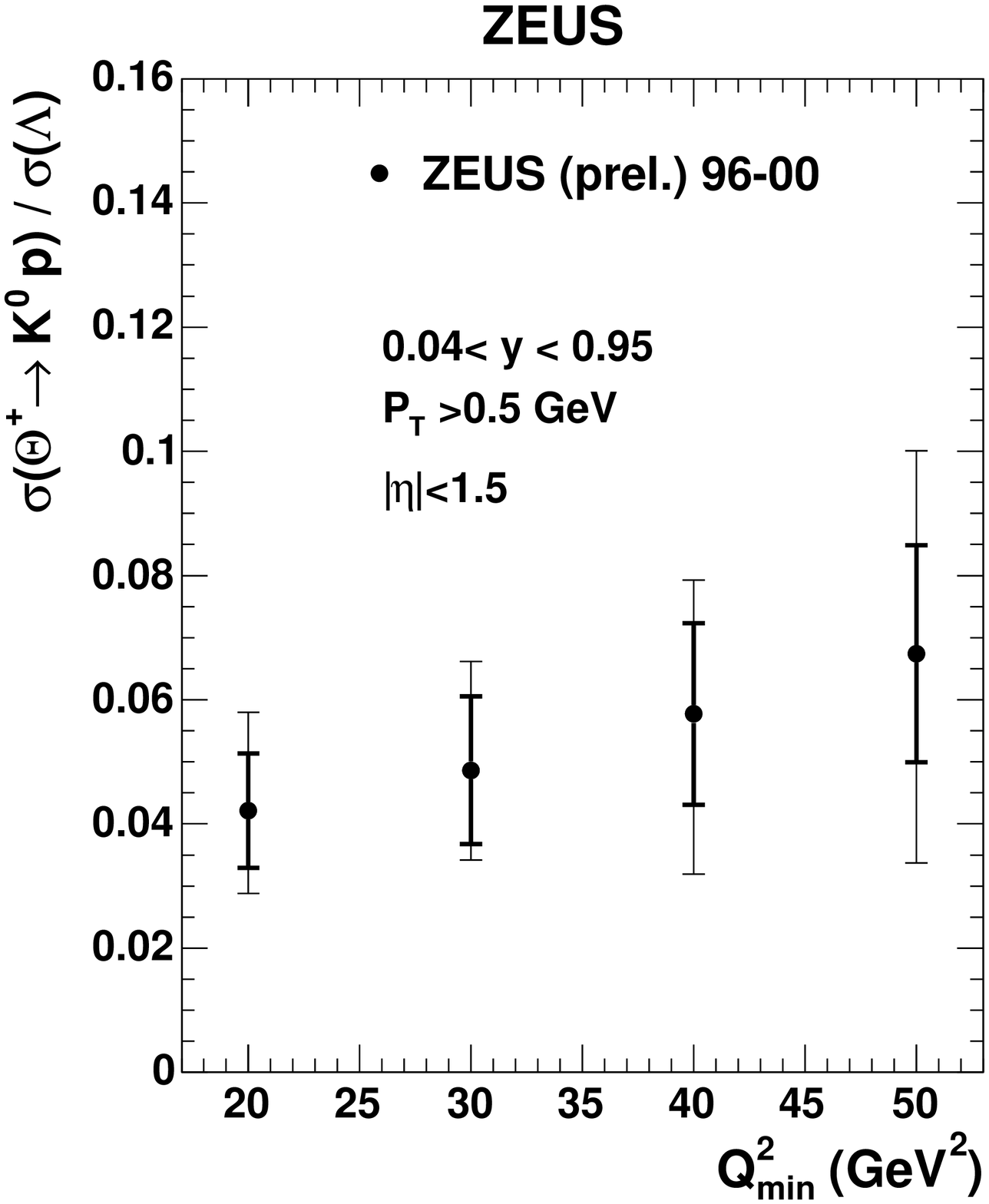}} 
 
 \vspace*{ -6.5cm}\hspace*{ 7.0cm}{        (b)}
 
 \vspace*{ +5.8cm}\hspace*{-2.0cm}                
 
\caption{ (a) Visible cross sections for the $\Theta^{\pm}$ baryon decaying to $K^0 p (\bar p)$
as a function of $Q^2_{min}$. (b) Cross section ratio                                 
$R = \sigma (\Theta^+\to K^0 p)/\sigma(\Lambda)$ as a function of $Q^2_{min}$.
            \label{Fig3}}
\end{figure}

The $\Theta$ production cross section was measured~\cite{ichep273} in the kinematic region
$Q^2 > 20$~GeV$^2$, $0.04 < y < 0.95$, $p_T (\Theta) > 0.5$~GeV and $|\eta (\Theta)|
< 1.5$  to be                                                                
  $\sigma (e p\to e \Theta^{\pm} X\to
                          e K^0 p^\pm X)=125\pm 27(stat.)^{+36}_{-28}(syst.) pb$,    
        where $y$ is the lepton inelasticity.                            The
acceptance was calculated using the RAPGAP MC, where $\Sigma^+$ baryons were treated
as $\Theta^+$         with $M=1.522$~GeV, forced to decay $100\%$ to
 $K^0_S p (\bar p)$. The                 $\Theta$ visible acceptance    was
$\approx 4\%$. Fig.~\ref{Fig3}(a) shows the cross section integrated above several
$Q^2_{min}$ values.                       The cross-section ratio         to that of 
    $\Lambda (1116)$,        
$R = \sigma (\Theta^+\to K^0 p)/\sigma(\Lambda)$ (antiparticles are included),
was measured in the same kinematic region. $\Lambda$ baryons were measured in the
decay mode $\Lambda\to p\pi^-$ and    protons were selected by $dE/dx$ with    
identical cuts as for the $\Theta^+$. The $\Lambda$ acceptance                
($\approx 10\%$)                                               was calculated
using the ARIADNE MC. The result for
$Q^2 > 20$~GeV$^2$ is $R=                                                           
          (4.2\pm 0.9^{+1.2}_{-0.9})\%$. Fig.~\ref{Fig3}(b) shows $R$ for these  
$Q^2_{min}$ values. It is                          not compatible with upper limits  
              from HERA-B and ALEPH, where $R < 0.5\%$.

 \vspace*{ -0.3cm}                                
If the observed $1.52$~GeV peak is due to a $I=1$ state, a $\Theta^{++}$ signal
is expected in the $K^+ p$ and $K^-\bar p$ spectrum. Selecting protons and charged
kaons using $dE/dx$, no peak was seen in the above distribution. A clean $10\sigma$
$\Lambda (1520)\to K^- p$ or $K^+\bar p$ signal was seen with mass and width  
consistent with the PDG values~\cite{PDG}. The number of $\Lambda (1520)$ and
$\bar\Lambda (1520)$ are similar.

 \vspace*{ -0.1cm}                                
\section{Search for pentaquarks in the $\Xi\pi$ channels}
 
The $pp$ fixed-target NA49 Collaboration ($\sqrt{s}=~17.2$~GeV)
  reported~\cite{NA49} observation of the $\Xi$ multiplet pentaquark candidates $\Xi^{--}_{3/2}$ and
   $\Xi^{0}_{3/2}$                                                               
                                                  predicted~\cite{Diakonov} at the bottom
of the anti-decuplet of baryons.  They found narrow peaks in these $\Xi\pi$ combinations
at $M\approx 1862$~MeV with a width $< 18$~MeV. The significance of the signal for the
sum of all 4 $\Xi\pi$ channels is $5.8\sigma$.
 
\begin{figure}[ht]
 
 \hspace*{-0.3cm}      
 \vspace*{-0.2cm}      
  \resizebox{14.3pc}{!}{\includegraphics{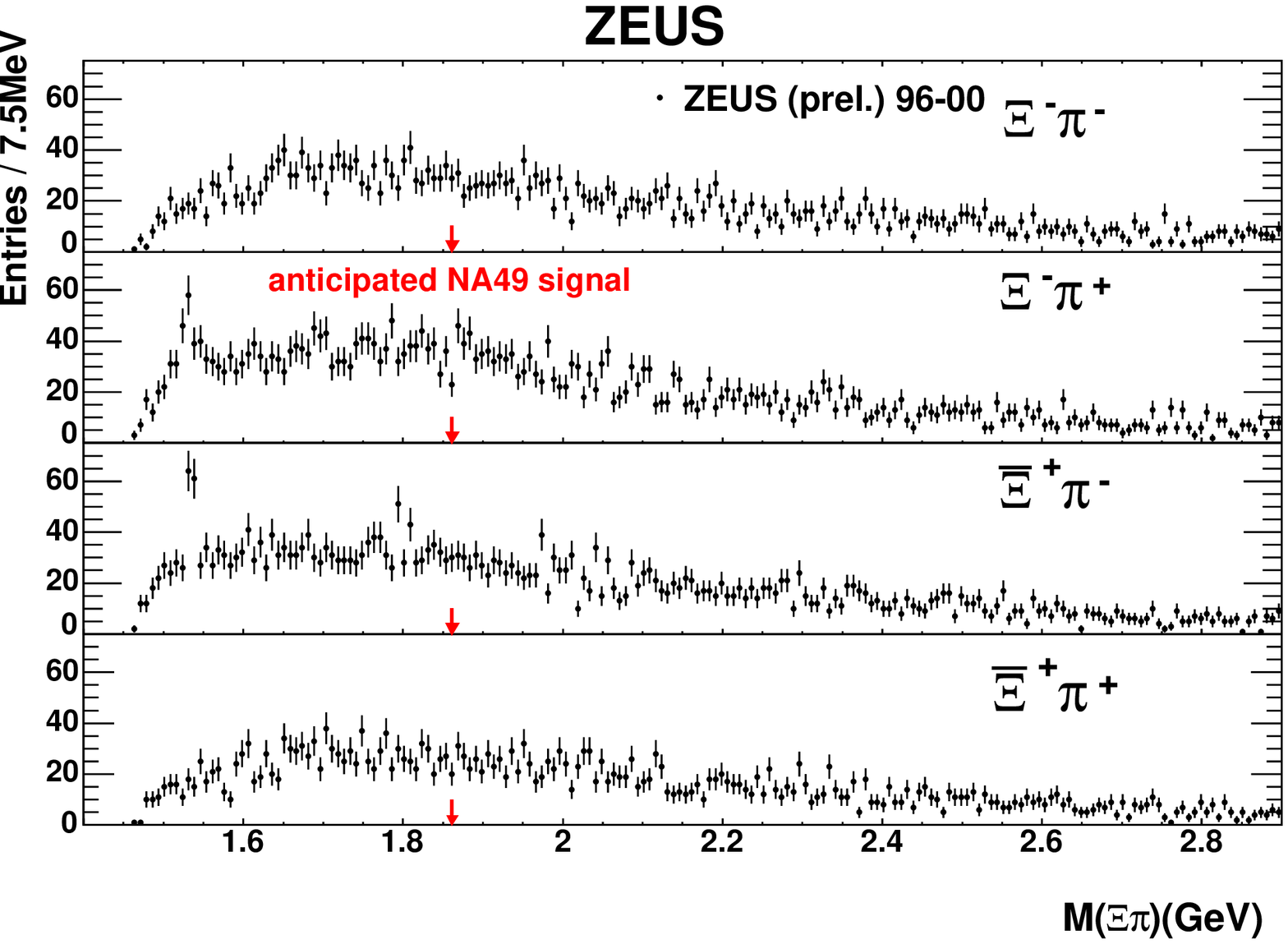}}

\vspace*{-4.4cm}\hspace*{+5.7cm}\resizebox{15.2pc}{!}{\includegraphics{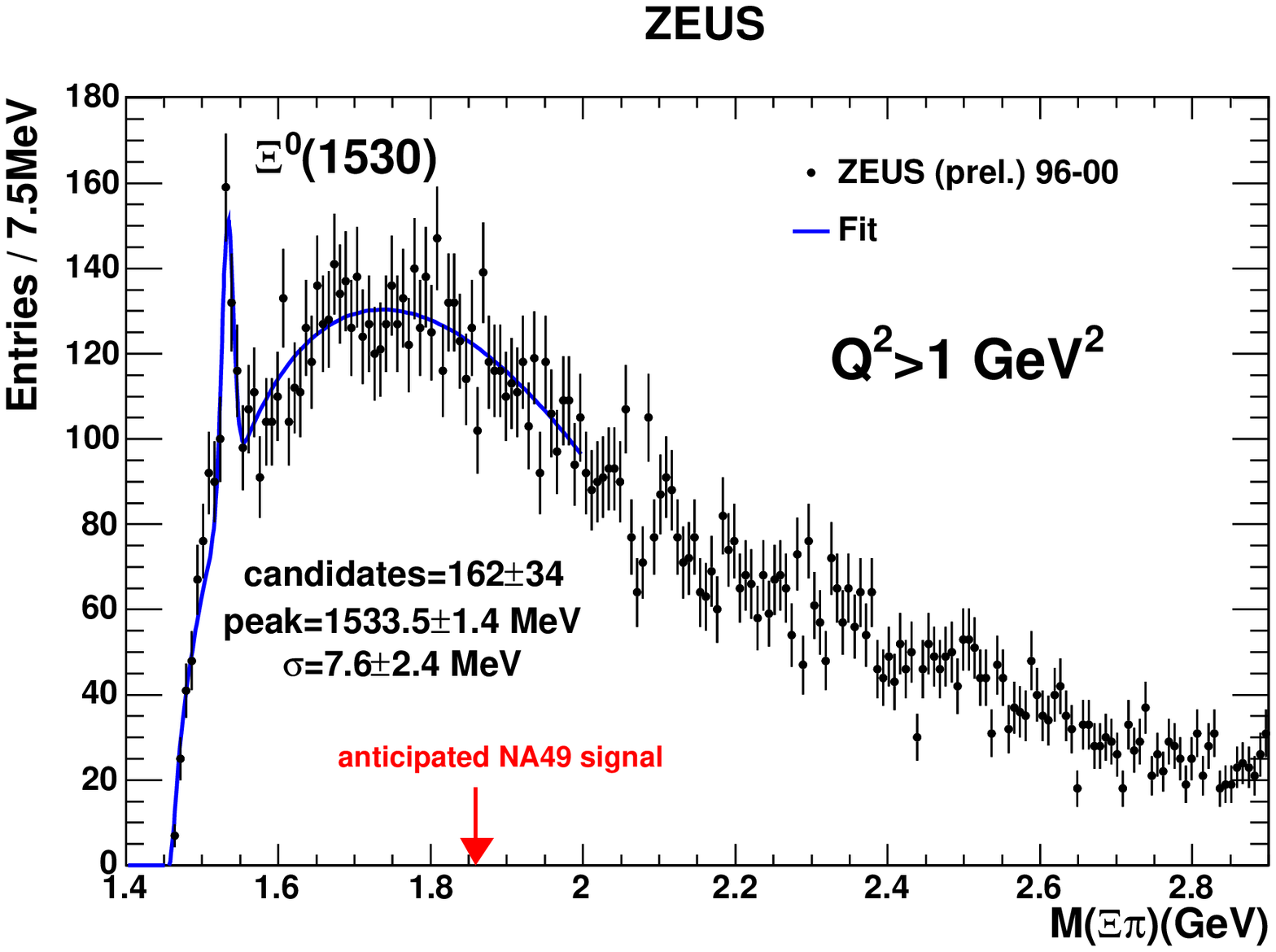}}

 \vspace*{+0.3cm}                      
\caption{$M(\Xi\pi)$ for $Q^2 > 1$~GeV$^2$ for each charge combination (left) 
and for all charge combinations combined (right).                               
            \label{Fig4}}
\end{figure}

ZEUS searched for such states in its DIS HERA-I data~\cite{ichep293}.
$\Xi^-$($\bar\Xi^+$) states were reconstructed via the $\Lambda\pi^-$($\bar\Lambda\pi^+$)
decay channel, with $\Lambda\to~p\pi^-~(\bar\Lambda\to~\bar{p}\pi^+$). Very clean $\Lambda$ 
and $\Xi$ signals were obtained with $\approx 130000~\Lambda +\bar\Lambda$ and 
$\approx 2600~\Xi + \bar\Xi$ candidates. In Fig.~\ref{Fig4} the $\Xi\pi$ invariant mass
spectrum for $Q^2 > 1$~GeV$^2$ is shown. The left histograms show each charge combination
separately. The right histogram is the sum of all $\Xi\pi$ combinations. A clean
$\Xi^0 (1530)\to \Xi\pi$ signal of $\approx 4.8\sigma$ is seen in the combined plot.
No evidence is seen for the NA49 signal around $1862$~MeV in any of the $\Xi\pi$ mass plots.
No signal is visible also with $Q^2 > 20$~GeV$^2$. The discrepancy may be due to the fact
that the ZEUS results come from the central rapidity region, while NA49 also covers the 
forward region.
 
 \vspace*{-0.2cm}      
 
 \vspace*{-0.8cm}                      
\section{Search for a charmed pentaquark decaying   to $D^{*\pm}p^{\mp}$}
 
The existence of the strange pentaquark $\Theta^+$ implies that charmed pentaquarks,
$\Theta^0_c =uudd\bar c$, should also exist. One type of model~\cite{Jaffe} predicts
$M(\Theta^0_c)\approx 2710$~MeV, which is below the threshold to decay strongly to $D$
mesons. Another model~\cite{Karliner} predicts a $\Theta^0_c$ which decays mainly
to $D^- p$ or $D^0 n$ (charge conjugate included) with
$M(\Theta^0_c) =  2985$~MeV and $\Gamma (\Theta^0_c )\approx~21$~MeV.         
If $M(\Theta^0_c)$ is above the sum of the $D^*$ and $p$ masses ($2948$~MeV), it can decay also
to $D^{*\pm}p^{\mp}$.
 
The H1 Collaboration found~\cite{H1} a narrow signal in the $D^{*\pm} p^{\mp}$ invariant mass at
$3.1$~GeV with a width consistent with the detector resolution. The signal was seen in a DIS
sample of $\approx 3400~D^{*\pm}\to D^0\pi^{\pm}\to (K^{\mp}\pi^{\pm})\pi^{\pm}$                   
       with a rate of $\approx 1\%$ of the visible $D^*$ production. A less clean signal  
of a comparable rate
was seen also in the H1 PHP sample.
 
\begin{figure}[ht]
 
 \vspace*{-0.6cm}      
  \resizebox{13.5pc}{!}{\includegraphics{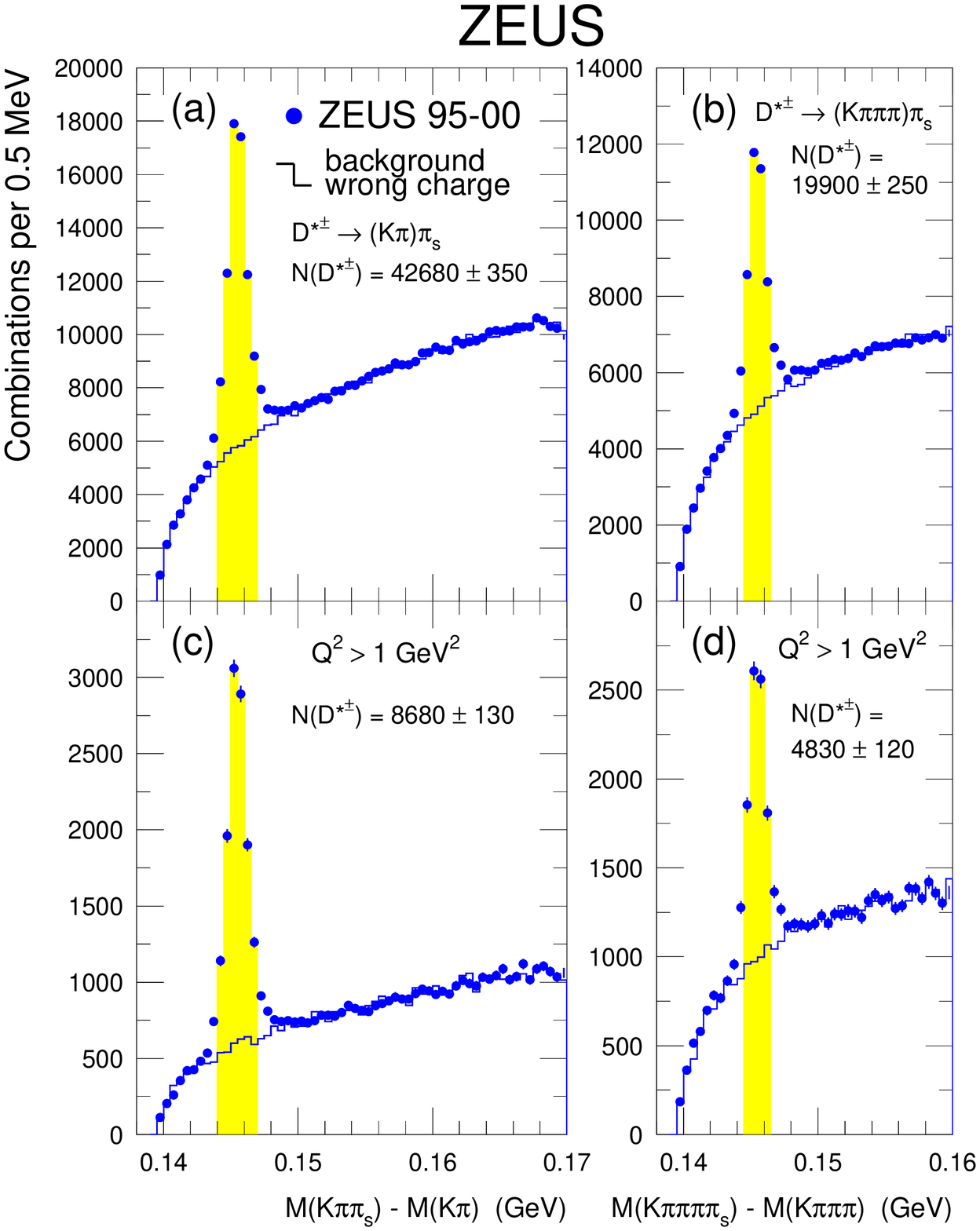}}
 
\vspace*{-7.8cm}\hspace*{+5.6cm}\resizebox{16.0pc}{!}{\includegraphics{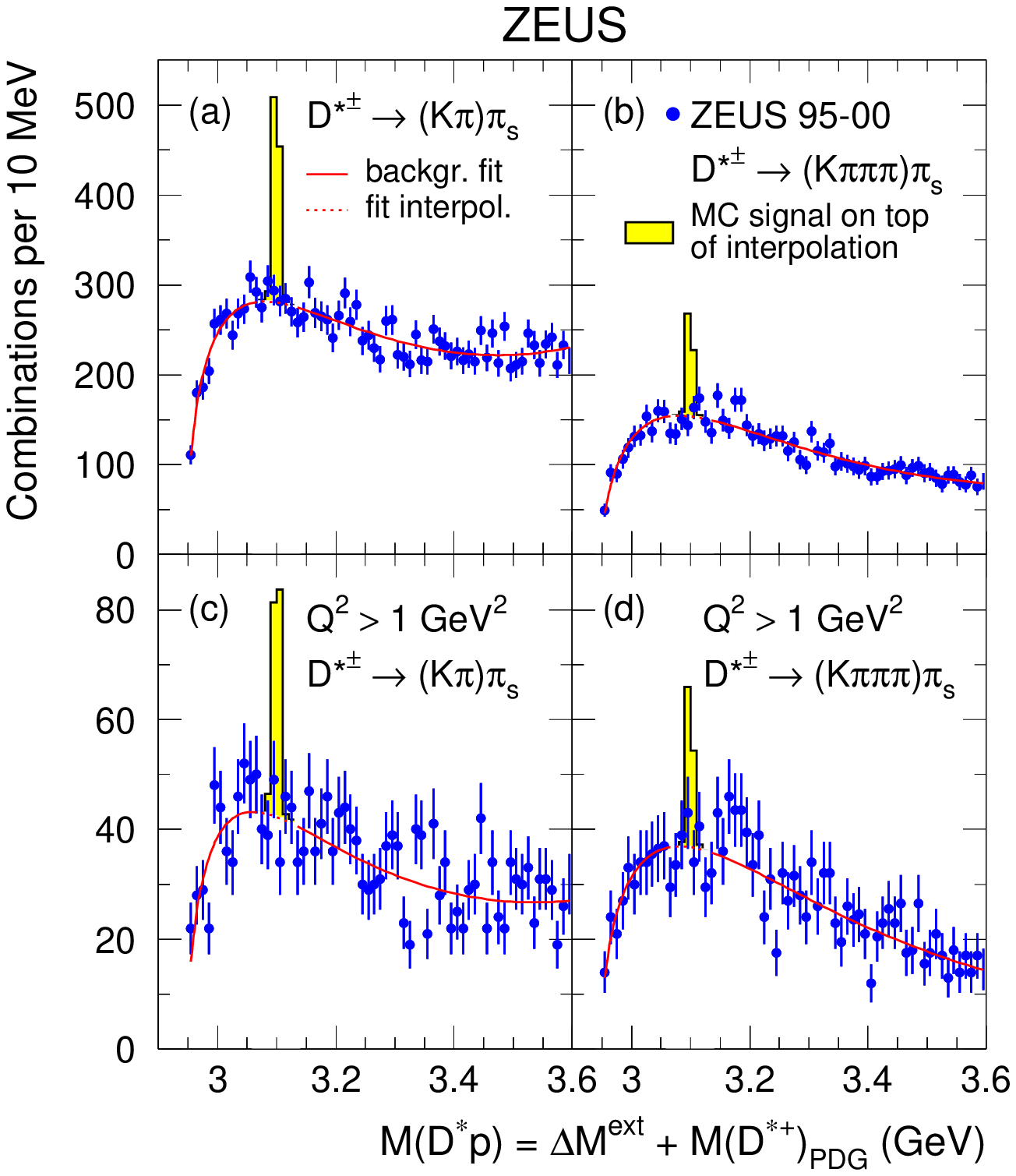}} 
 
 \vspace*{+0.1cm}      
\caption{Left:~$\Delta M$ distributions (dots) for (a) $D^*\to K\pi\pi$ and (b) $D^*\to K\pi\pi\pi\pi$
candidates. Events with $Q^2 > 1$~GeV$^2$ for the two channels, respectively, are shown          
in (c) and (d). The histograms are for wrong charge combinations.                  
Right:~$M(D^{*\pm}p^{\mp})$ distributions (dots) for the same samples. Solid curves are fits
to a background function (see text). Shaded historgams are MC $\Theta^0_c$ signals, normalised to 
$\Theta^0_c /D^* =1\%$, on top of the background fit.
            \label{Fig5}}
\end{figure}

The $\Theta^0_c$ search of ZEUS in the $D^{*\pm} p^{\mp}$
                                               mode was performed with the full HERA-I data~\cite{thetac}.
Clean $D^{*\pm}$ signals were seen in the $\Delta M~=~M(D^{*\pm})-~M(D^0)$ plots (Fig.~\ref{Fig5} left).
Two $D^{*\pm}\to~D^0\pi^{\pm}$      
          decay channels were used with
 $D^0\to~K^{\mp}\pi^{\pm}$ and               
 $D^0\to K^{\mp}\pi^{\pm}\pi^+\pi^-$.               
The $\Theta^0_c$ search was performed in the kinematic range                                         
          $|\eta (D^*)|< 1.6$ and                            
          $p_T~(D^*)~>~1.35~(2.8)$~GeV and with $\Delta M$ values between
   $0.144~-~0.147 (0.1445 - 0.1465)$~GeV for the 
        $K\pi\pi$ ($K\pi\pi\pi\pi$) channel.                                                      
                         In these shaded bands a total        of                                     
         $\approx 62000~D^*$'s was obtained (Fig.~\ref{Fig5}a-b left)                         
                                             after subtracting wrong-charge combinations
with       charge $\pm 2$ for the $D^0$ candidate.                                     
Selecting DIS events with $Q^2 > 1$~GeV$^2$ yielded smaller, but cleaner $D^*$ signals with a
total of $\approx 13500~D^*$'s (Fig.~\ref{Fig5}c-d left).
                
 
\begin{figure}[ht]

 \vspace*{-0.5cm}      
 \hspace*{-0.2cm}      
  \resizebox{15.4pc}{!}{\includegraphics{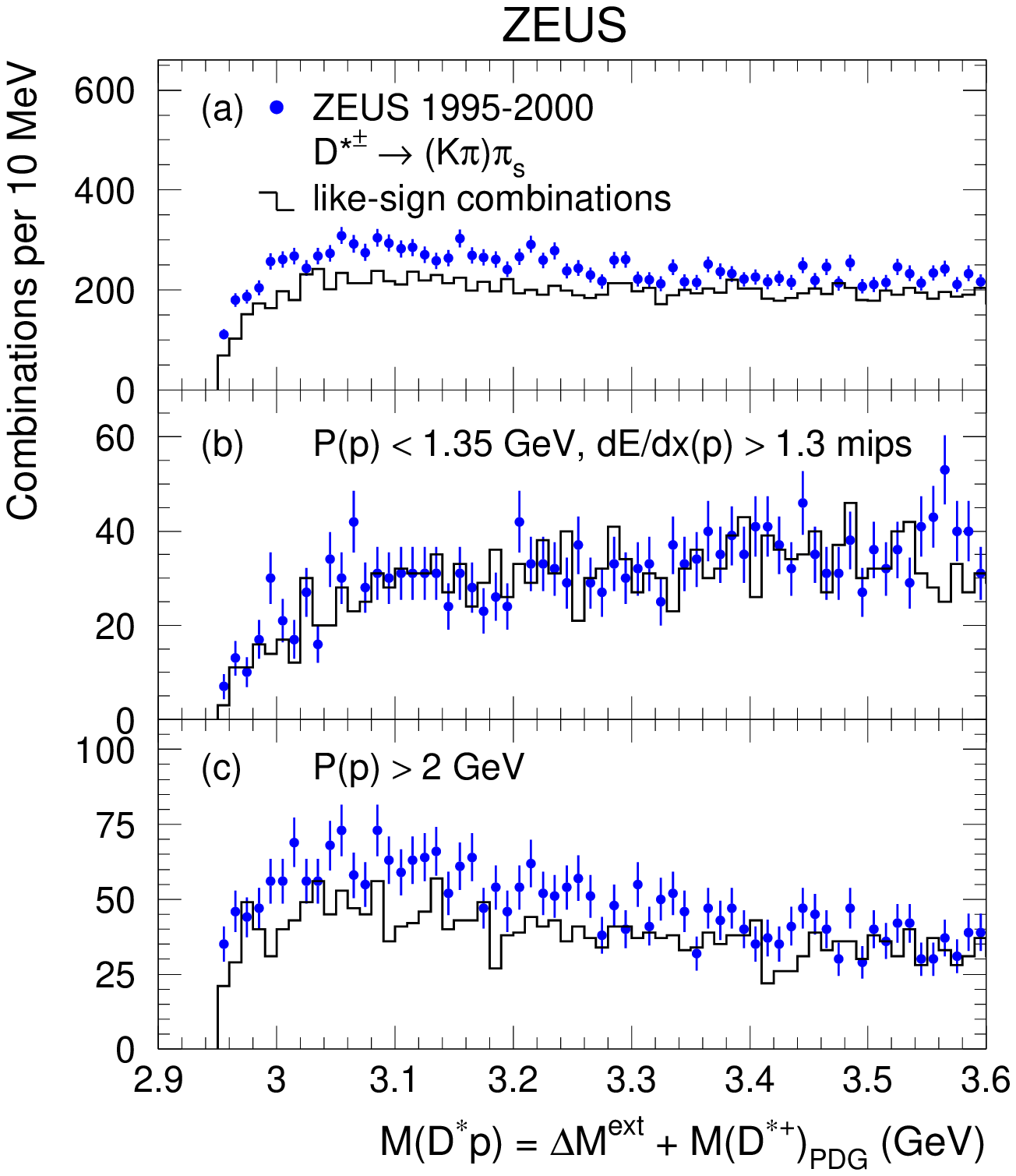}}
 

\vspace*{-7.3cm}\hspace*{+5.6cm}\resizebox{15.4pc}{!}{\includegraphics{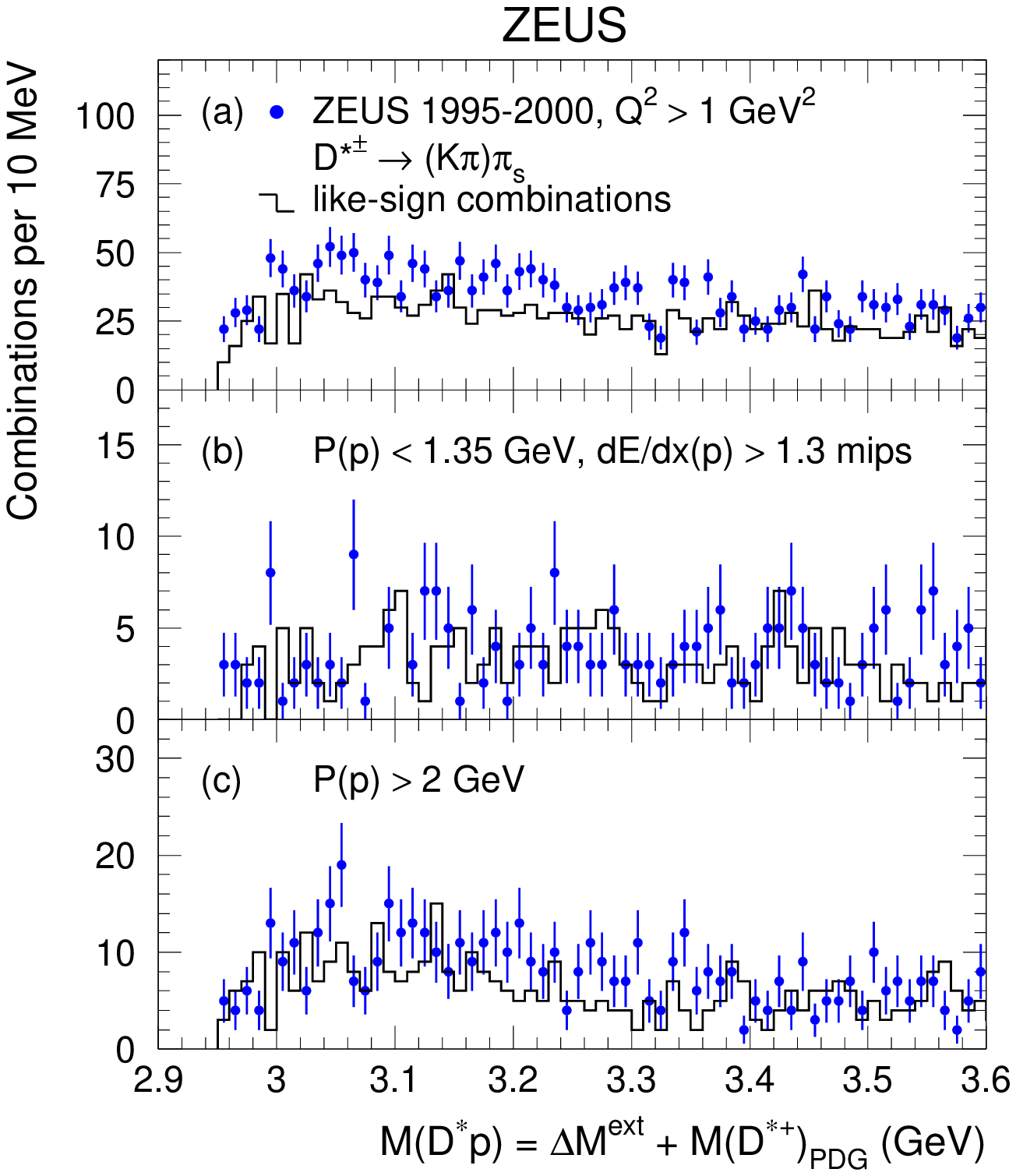}}

\vspace*{-0.2cm}                                                                          
\caption{Left:~$M(D^{*\pm}p^{\mp})$ distributions for the $K\pi\pi$ channel                  
                                                  (dots) with (a) all proton candidates, (b) 
candidates with $P(p) < 1.35$~GeV and $dE/dx > 1.3$, and (c) candidates with $P(p) > 2$~GeV.
Histograms show the $M(D^{*\pm}p^{\pm})$ like-sign combinations. Right: Same for DIS events
with $Q^2 > 1$~GeV$^2$.
            \label{Fig6}}
\end{figure}

Protons were selected with $p_T (p) > 0.15$~GeV. To reduce the pion and kaon background,
a   parameterisation of the expected $dE/dx$ as a function of $P/m$ was obtained using
tagged protons from $\Lambda$ decays and tagged pions from $K^0_S$ decays. The $\chi^2$
probability of the proton hypothesis was required to be above $0.15$. 
Fig.~\ref{Fig6} shows the $M(D^* p) = M(K\pi\pi p) - M(K\pi\pi) + M(D^*)_{PDG}$          
                                     distributions for the $K\pi\pi$ channel for the full
(left) and the DIS (right) samples, where $M(D^*)_{PDG}$ is the $D^{*\pm}$ mass~\cite{PDG}.
                                    In the low-$P$ selection (Fig.~\ref{Fig6}b), a
clean proton sample separated from the $\pi$ and $K$ $dE/dx$ bands was obtained by taking only
tracks with $P < 1.35$~GeV and $dE/dx > 1.3$~mips. In the high-$P$ selection (Fig.~\ref{Fig6}c)
                                                                             only tracks
with $P(p) > 2$~GeV were used. The latter selection was prompted by the H1 observation~\cite{H1}
of a better $\Theta^0_c$ signal-to-background ratio for high proton momenta. No narrow
signal is seen in the $K\pi\pi$ (Fig.~\ref{Fig6}) as well as in the                          
        $K\pi\pi\pi\pi$ (Fig.~\ref{Fig5}b,d right) channel.
The $K\pi\pi$ analysis was repeated               using very similar selection
criteria as   in the H1 analysis~\cite{H1}. No indication of a narrow resonance was found
in either the DIS or the PHP event sample~\cite{thetac}.
 
 
$95\%$ C.L. upper limits on the fraction of $D^*$ mesons originating from  $\Theta^0_c$ decays,               
$R(\Theta^0_c\to D^* p/D^*)$,             were calculated
in a signal window $3.07~<~M(D^* p)~<~3.13$~GeV for the $K\pi\pi$ and $K\pi\pi\pi\pi$ channels.
A visible rate of $1\%$ for this fraction (Fig.~\ref{Fig5} right),
as claimed by H1~\cite{H1},                                           is excluded by
$9\sigma$ ($5\sigma$) for the full (DIS) combined sample.
                         The $M(D^* p)$ distributions were fitted                           to
 the form $x^a e^{-bx+cx^2}$, where $x=M(D^* p)-M(D^*)-m_p$ (Fig.~\ref{Fig5} right).     
The number of reconstructed $\Theta^0_c$ baryons was estimated by subtracting in the signal window
the background function from the observed number of events, yielding        
$R(\Theta^0_c\to D^* p/D^*) < 0.23\%$ and $ < 0.35\%$ for the full and DIS combined two channels.
The acceptance-corrected rates are, respectively, $0.37\%$ and $0.51\%$.
The $95\%$ C.L. upper limit  on the fraction of charm quarks fragmenting to $\Theta^0_c$ times the
branching ratio    $\Theta^0_c\to D^* p$ for the combined two channels is
$f(c\to\Theta^0_c)\cdot B_{\Theta^0_c\to~D^* p} < 0.16\%$ ($ < 0.19\%$) for the full (DIS) sample.

\vspace*{-0.2cm}                                                                          
\section{Summary}
 
The ZEUS HERA-I data sample was used to search for narrow baryonic pentaquark candidates.              
For the inclusive DIS sample  a $4.6\sigma$ narrow
signal was seen in the fragmentation region in the combined $M(K^0_S p)$ and $M(K^0_S\bar p)$ plot    
          at the $\Theta^{\pm}$ mass range.
If due to the $\Theta$ baryon, this is the first evidence for the anti-pentaquark $\Theta^-$.
The cross-section ratio        
$\sigma (\Theta^+\to K^0 p)/\sigma(\Lambda)$ for $Q^2 > 20$~GeV$^2$ is                
         $(4.2\pm 0.9^{+1.2}_{-0.9})\%$.  No evidence is found for the NA49 $\Xi\pi$ signal at
$1862$~MeV in the inclusive DIS sample. No resonance structure is seen in                             
    $M(D^{*\pm}p^{\mp})$ around $3.1$~GeV.
The $95\%$ C.L. upper limit on the visible rate
$R(\Theta^0_c\to D^* p/D^*)$ is $0.23\%$ ($0.35\%$ for DIS). The ZEUS data are not compatible
with the H1 result of $\approx 1\%$ of the above rate. Such a rate is excluded by $9\sigma$ for
the full data and by $5\sigma$ for the ZEUS DIS data.


\vspace*{-0.2cm}                                                                          
\begin{thebibliography}{0}
\bibitem{Diakonov} D. Diakonov, V. Petrov and M.V. Polyakov, {\it Z. Phys. }
{\bf A359}, 305 (1997).
 
\bibitem{zeustheta} ZEUS Coll., S. Chekanov et al., {\it Phys. Lett.}  
{\bf B591}, 7 (2004).
 
\bibitem{PDG} Particle Data Group, K. Hagiwara et al., {\it Phys. Rev.} {\bf D66}, 
10001 (2002).
 
\bibitem{ichep273} ZEUS Coll., Abstract 273, XXXII Int. Conf. on High Energy Physics, ICHEP2004,
August 2004, Beijing, China.               
 
\bibitem{NA49} NA49 Coll., C. Alt et al.,                   , 
{\it Phys. Rev. Lett.} {\bf 92}, 42003 (2004).
 
\bibitem{ichep293} ZEUS Coll., Abstract 293, XXXII Int. Conf. on High Energy Physics, ICHEP2004,
August 2004, Beijing, China.               
 
\bibitem{Jaffe} A. Jaffe and F. Wilczek,                     , 
{\it Phys. Rev. Lett.} {\bf 91}, 232003 (2003).
 
\bibitem{Karliner} M. Karliner and H.J. Lipkin, Preprint hep-ph/0307343 (2003).     
 
\bibitem{H1} H1 Coll., C. Atkas et al.,                   , 
{\it Phys. Lett.} {\bf B588}, 17 (2004). 
 
\bibitem{thetac} ZEUS Coll., S. Chekanov et al., hep-ex/0409033,              
            Eur. Phys. J.,       in print.
 
\end{thebibliography}
\end{document}